\documentclass[conference]{IEEEtran}
\usepackage{rotating}
\usepackage[autostyle]{csquotes}
\usepackage[hyphens]{url}
\usepackage[colorlinks=true, linkcolor=cyan, urlcolor=cyan, citecolor=cyan]{hyperref}
\usepackage{amsmath,amssymb,amsfonts}
\usepackage{graphicx}
\usepackage{xcolor}
\usepackage{caption}
\usepackage{subcaption}
\usepackage{balance}
\usepackage[inline]{enumitem}
\usepackage{soul}
\usepackage{setspace}
\usepackage{cite}

\newcommand{\mydot}[1][black]{{\Large\textcolor{#1}{\ensuremath\bullet}} }
\newcommand{\mycross}[1][black]{{\Large\textcolor{#1}{\ensuremath\times}} }

\newcommand{\hide}[1]{}

\title{A Multi-Functional Web Tool for Comprehensive Threat Detection Through IP Address Analysis}

\author{Cebajel Tanan$^*$, Sameer G. Kulkarni$^\S$\textsuperscript{1}, Tamal Das$^*$ and Manjesh K. Hanawal$^\ddag$\textsuperscript{2} \\~\\

$^*$\textit{Indian Institute of Technology, Dharwad, India}\\
$^\S$\textit{Indian Institute of Technology, Gandhinagar, India}\\
$^\ddag$\textit{Indian Institute of Technology, Bombay, India}\\
$^*$\{210010055, tamal\}@iitdh.ac.in, $^\S$sameergk@iitgn.ac.in, $^\ddag$mhanawal@iitb.ac.in
}

\begin{document}

\maketitle

\addtocounter{footnote}{1}
\footnotetext{Sameer G. Kulkarni acknowledges funding support from SERB, Govt. of India, through the core research grant: CRG/2023/008021 and the Department of Telecommunications, Govt. of India, 5G Use Case Labs.}

\addtocounter{footnote}{1}
\footnotetext{Manjesh Kumar Hanawal acknowledges funding support from SERB, Govt. of India, through the Core Research Grant (CRG/2022/008807) and the Bureau of Police Research and Development (BPR\&D).
}
% \footnotetext{Department of Computer Science and Engineering, Indian Institute of Technology Dharwad. Email:\href{mailto:210010055@iitdh.ac.in}{210010055@iitdh.ac.in}}

% \addtocounter{footnote}{1}
% \footnotetext{Department of Computer Science and Engineering, Indian Institute of Technology Gandhinagar. Email:\href{mailto:sameergk@iitgn.ac.in}{sameergk@iitgn.ac.in}}

% \addtocounter{footnote}{1}
% \footnotetext{Department of Computer Science and Engineering, Indian Institute of Technology Dharwad. Email:\href{mailto:tamal@iitdh.ac.in}{tamal@iitdh.ac.in}}

% \addtocounter{footnote}{1}
% \footnotetext{Department of Computer Science and Engineering, Indian Institute of Technology Bombay. Email:\href{mailto:mhanawal@iitb.ac.in}{mhanawal@iitb.ac.in}}

\begin{abstract}
In recent years, the advances in digitalisation have also adversely contributed to the significant rise in cybercrimes. Hence, building the threat intelligence to shield against rising cybercrimes has become a fundamental requisite. 
Internet Protocol (IP) addresses play a crucial role in the threat intelligence and prevention of cyber crimes. 
However, we have noticed the lack of one-stop, free, and open-source tools that can analyse IP addresses. 
Hence, this work introduces a comprehensive web tool for advanced IP address characterisation.
Our tool offers a wide range of features, including geolocation, blocklist check, VPN detection, proxy detection, bot detection, Tor detection, port scan, and accurate domain statistics that include the details about the name servers and registrar information.

In addition, our tool calculates a confidence score %for each query 
based on a weighted sum of publicly accessible online results from different reliable sources to give users a dependable measure of accuracy. 
Further, to improve performance, our tool also incorporates a local database for caching the results, to enable fast content retrieval with minimal external Web API calls. Our tool supports domain names and IPv4 addresses, making it a multi-functional and powerful IP analyser tool for threat intelligence. Our tool is available at www.ipanalyzer.in

\end{abstract}


\section{Introduction}

Cybersecurity is a vast field that encompasses strategies for protecting digital assets and information. It includes data collection techniques, network encryption, malicious activity detection, cybercrime analysis, etc. The goal of cyber security is to protect against a wide variety of computer threats, including malware, ransomware, phishing, hacking, data breaches, and various cyber-attacks Analysis, Threat Detection and Response, Encryption, Access control, identity management, and security awareness training. In an increasingly digitized world, where cyber threats continue to evolve, effective security measures are needed to protect individuals, organizations, governments and critical infrastructure from potential injury.

In today’s world, cybersecurity is critical in maintaining our data and continuing core business operations and tasks, providing protection against computer systems, networks, and stored data that hackers could steal, compromise, or intercept. Strong cybersecurity measures are needed to protect data privacy, prevent financial loss or reputational damage and prevent business disruptions. This collective action is to combat ever-changing threats aimed at combining clearly defined technical tools and methods with professional efforts.

An Internet Protocol (IP) address is a number assigned to each device connected to a computer network that uses the Internet Protocol to communicate. IP addresses serve two main functions: host or network interface identification and location addressing. There are two main types of IP addresses in use today: IPv4 (Internet Protocol version 4) and IPv6 (Internet Protocol version 6). IPv4 addresses consist of four numbers separated by dots, such as \texttt{192.168.0.1}, while IPv6 addresses are long and encoded in hexadecimal, such as \texttt{2001:0db8:85a3:0000:0000:8a2e:0370:7334}. IP addresses serve as the digital addresses for routing the data packets to the correct device bearing that IP address.

IP address analysis refers to collecting and assessing relevant details on a particular IP address to explain its source or origin, trustworthiness, risks, activities, and so forth. This procedure is commonplace in computer security, network administration, and electronic investigations, which are focused on establishing the presence of threats, protecting systems from them, or simply understanding where the traffic to a particular resource comes from. Aspects of IP Analysis, like geolocation, proxy check, virtual private network (VPN) check, blocklist check, etc., are critical for varied reasons that we elaborate in the next section (\S\ref{sec:bkgd}).

The rationale behind collecting such data is to form a complete picture of the specific IP address, which should guide the decision whether to permit, deny, or track traffic associated with that IP address. In this regard, IP analysis provides the impetus for preventing attacks, managing threat situations, and protecting the organisation’s network against illicit access in cybersecurity teams. By analysing IP addresses, cybersecurity professionals can identify potential threats such as malware, phishing attempts, and hacking attempts. Traffic system monitoring enables the detection of anomalies, such as unusually high activity or connections to known bad areas, indicating potential security breaches. Geolocation data derived from IP addresses helps locate the source of the attack, and restrictions are imposed based on the access point. Integrating threat notification allows you to stay ahead of emerging threats and known malicious IPs, increasing security protection. Analysing IP addresses during incident response efforts helps reconfigure incidents, identify threats that lead to attacks, and assess the scope of a contract. Monitoring IP Data helps analyse incoming reliable connections so organisations can block or disable malicious IP addresses. It is also helpful in the area of digital forensics, where it helps in the identification of offenders and their strategies.

Some of the various fields where IP Analysis is critical are mentioned below:
\begin{enumerate}
    \item \textit{Threat detection and prevention:} IP reputation databases track the history of IP addresses, identifying those associated with malicious activity like spam or malware distribution. Analysing incoming traffic and blocking connections from flagged IPs helps prevent cyberattacks.

    \item \textit{Investigating cybercrimes:} After a security breach, analysing the source IP address can provide a starting point for tracing and potentially identifying the attacker's location.

    \item \textit{Law Enforcement:} Analysing IP addresses can help link individuals to specific online activities, aiding investigations into cybercrime, online harassment, or other illegal online activities. With the correct tools, we can identify whether individuals use some services to hide their identities, like Tor (onion routing project) or VPN (a virtual private network).

\end{enumerate}

We have surveyed several online tools available in the market for IP Analysis for Geolocation, Proxy Detection, VPN Detection, Tor Detection, Bot Detection, Threat Detection, Port Scan, Liveness Test, and Whois Information \cite{survey}. Out of the 105 tools analysed, we observed that none provided a comprehensive set of features. To address this problem, we are introducing a web tool, a one-stop solution for comprehensive IP address analysis, as it incorporates critical features for IP address analysis, which we will discuss in the next section. On top of that, our tool also provides robust industry-standard user management. We will elaborate on the features of our web tool in the proposed solution Section~\ref{sec:proposed}.

% We can stop various attacks and possible disasters by carefully analysing IP addresses. Some famous examples include:
% \begin{enumerate*}[label=(\arabic*)]
%     \item \textit{Data Breaches:} While not solely due to a lack of IP analysis, major data breaches like the Equifax breach \cite{equifax}\cite{equifax2} in 2017 (affecting millions of individuals) involved attackers potentially leveraging anonymization tools or compromised infrastructure, making it challenging to pinpoint their origin solely through IP analysis.

%     \item \textit{Cyberattacks on Critical Infrastructure Incidents} like the 2020 SolarWinds supply chain attack \cite{solar}, where attackers compromised a software company's infrastructure, highlight the evolving nature of cyber threats. While IP analysis might identify the initial point of compromise, the attackers often use sophisticated techniques to mask their origin and exploit vulnerabilities within trusted systems, making it a complex investigative process.

%     \item \textit{The 2016 Bangladesh Bank Heist:} \cite{bank} Hackers infiltrated the bank's network and attempted to steal nearly \$1 billion through fraudulent transfers. While some reports suggest the attackers used anonymization tools and originated from outside Bangladesh, IP analysis, combined with other investigative techniques like tracing financial transactions, eventually helped identify individuals involved in the heist. This case demonstrates the potential value of IP analysis when used collaboratively with other investigative methods.
    
% \end{enumerate*}

\section{Background}\label{sec:bkgd}
IP Analysis helps to find several characteristic features of an IP address. In this section, we describe the definitions of these features along with the corresponding methodologies to find them (in Section~\ref{features}), as well as the various tools used for IP address analysis (in Section~\ref{tools}).

\subsection{Features} \label{features}

% IP Geolocation
\subsubsection{\textbf{IP geolocation}} This method attempts to determine the geographical location in the form of (country, city) associated with an IP address \cite{maxmind}.

\subsubsection*{Methodology} Using good databases periodically maps IP addresses to their approximate locations \cite{tax1, maxmind, geolocation-limitations, geolocation-privacy}.

\subsubsection*{Accuracy Concerns} Geolocation databases might not always be accurate, especially for mobile devices or due to the anonymization services. Accuracy also depends on the freshness of the information in the database. However, the database update period varies from database to database, typically from 24 hours to a couple of weeks. Some databases, which charge some premium, offer more recent information by updating every minute, whereas community databases are not updated that regularly \cite{geolocation-limitations, maxmind-accuracy, geolocation-accuracy2}.

\subsubsection*{Privacy Concerns} Collecting and using geolocation data raises privacy concerns, and hence requires adherence to the relevant regulations~\cite{geolocation-privacy}.

% Tor Detection
\subsubsection{\textbf{Tor Detection}} Tor (The Onion Router) is a privacy-focused network that anonymizes users' Internet traffic by routing it through multiple layers of encryption and volunteer-run relays around the world. Tor detection is the process of identifying traffic originating from the Tor network \cite{tor-defn, tor-method}.

\subsubsection*{Methodology}
A few relevant methodologies in this regard are as follows.
\begin{itemize}
    \item \textit{Public Tor Exit Nodes:} Tor publishes a list of ``exit nodes" -- the final points from which Tor traffic enters the regular Internet. Organizations can compare incoming IP addresses against this list to identify Tor connections \cite{tor-method, tax5, tor-method2, tor-method3}.

    \item \textit{Browser Fingerprinting:} Using browser fingerprinting methods, one can check for distinct features of the Tor browser. Example: Tor Browser returns a blank image (RGB 255, 255, 255) when extracting the base64-encoded value of an HTML5 canvas element \cite{tax5}.
\end{itemize}

\subsubsection*{Accuracy Concerns} The accuracy of Tor detection depends on the methods used and how up-to-date the information is. Detecting listed exit nodes is highly accurate but only catches a portion of overall Tor traffic. Statistical analysis or behavioural fingerprinting can detect Tor traffic even without an exit node list. These methods are less accurate and prone to more false positives \cite{tor-accuracy}.

\subsubsection*{Limitations} Not all Tor traffic can be detected as users can configure Tor to use unlisted ``bridges" \cite{tor-bridge} or obfuscate their traffic, making detection more difficult. While Tor detection can identify traffic belonging to the Tor network, it does not reveal the specific user behind the connection. Some legitimate services or anonymity tools might utilize exit nodes used by Tor, leading to potential false positives \cite{tor-accuracy}.

% VPN Detection
\subsubsection{\textbf{VPN Detection}}
VPNs (Virtual Private Networks) encrypt a user's Internet traffic and route it through intermediary servers, masking their actual IP address and potentially circumventing geographic restrictions. VPN detection is the process of identifying traffic coming from VPN services \cite{tax6}.

\subsubsection*{Methodology} A few relevant methodologies in this regard are as follows.
            \begin{itemize}
                \item \textit{Known VPN IP Addresses:} Services maintain databases of IP addresses associated with commercial VPN providers. Comparing incoming traffic to these lists can identify the usage of VPN \cite{tax6}.

                \item \textit{Port Scanning:} VPNs often use specific ports for communication. Scanning for open ports associated with standard VPN protocols can indicate VPN usage \cite{tax6}.

                \item \textit{Deep Packet Inspection (DPI):} Analysing the contents of network packets can reveal patterns or signatures that suggest VPN traffic, even if the IP address itself is not recognised \cite{tax6, vpn-method}.

                \item \textit{TCP/IP Fingerprinting:} Comparing the Operating System (OS) induced from the TCP/IP fingerprint with the OS advertised by the User-Agent. From analysing the TCP/IP packets exchanged between the device and the web server, we can discover the device's operating system, which indicates an OS mismatch compared to the browser's operating system. Inconsistencies detected in this case may indicate the presence of a VPN, proxy server, or Apple's iCloud Private Relay \cite{tax6}.
            \end{itemize}

\subsubsection*{Accuracy Concerns} Basic IP-based detection can be highly accurate for well-known commercial VPNs but less effective for custom or lesser-known VPN providers. Deep Packet Inspection (DPI) and behavioural analysis offer a higher detection rate, especially for sophisticated techniques, but may have a higher likelihood of false positives \cite{tax6}.
                
\subsubsection*{Limitations} VPN providers actively change tactics to avoid detection, making it an ongoing challenge. Legitimate services might use techniques similar to VPNs, leading to potential false positives in detection. Some VPNs offer obfuscation options to mask traffic, making detection even more challenging, \textit{e.g.}, changing header fields in packets to simulate non-VPN-like behaviour. Also, port scanning is not straightforward, as a firewall can drop packets sent to the device. The same goes for identifying the device's OS using custom packets apart from TCP packets sent to the device's browser.

% Proxy Detection
\subsubsection{\textbf{Proxy Detection}} 
Proxy detection is the process of identifying Internet traffic originating from a proxy server. Proxy servers act as intermediaries between a user's device and the Internet, masking the user's IP address and sometimes location \cite{proxy}.

\subsubsection*{Methodology} A few relevant methodologies in this regard are as follows. \begin{itemize} 
            \item \textit{Latency Analysis:} Compare latency from browser to server with the latency from a web server to an external IP address. If both latency measurements differ significantly (namely, the Browser to Server Latency is significantly higher than the Server to Browser Latency), it is possible to conjecture that there is an intermediate host between the browser and the web server \cite{proxy-method, tax7}.

            \item \textit{Port Scanning:} Specific ports are often associated with proxy communication protocols. Scanning for open ports used by known proxy protocols can indicate their use \cite{tax7}.

            \item \textit{WebRTC Leaks:} WebRTC (Web Real-Time Communication) is a technology that can inadvertently reveal a user's actual IP address even when using a proxy if not configured correctly \cite{proxy-method, tax7}.

            \item \textit{DNS Leaks:} Network configurations on the user device can lead to DNS (Domain Name System) requests that can sometimes leak the user's actual IP address, even when they are using a proxy \cite{proxy-method, tax7, DNSLeakTest}. %depending on the configuration.

            \item \textit{TCP/IP Fingerprinting:} Compare the OS induced from the TCP/IP fingerprint with the OS advertised by the User-Agent. From analysing the TCP/IP packets exchanged between the device and the web server, we can discover the device's operating system, which indicates an OS mismatch compared to the browser's operating system. Inconsistencies detected in this case may indicate the presence of a VPN, proxy server, or Apple's iCloud Private Relay \cite{proxy-method, tax7}.

            \item \textit{Datacenter IP:} Check if the IP address belongs to a datacenter. Datacenter proxies are often hosted in public data centres like AWS or Digitalocean. Those cloud providers publish their IP ranges, making it possible to check whether a proxy belongs to a data center \cite{proxy-method, tax7}.

            \item \textit{HTTP Proxy Headers Analysis:} Many HTTP proxy servers add additional HTTP headers to each HTTP request. The presence of those headers indicates that a proxy server is used \cite{proxy-method, tax7}.
            
            \end{itemize}
            
\subsubsection*{Accuracy Concerns} Simple database checking can be practical for identifying basic proxies but less effective for more advanced ones. Combining techniques like port scanning, traffic analysis, and leak detection can improve accuracy but may still have limitations \cite{proxy-accuracy}.

\subsubsection*{Limitations} Datacenter-based detection only catches known proxies. New proxies emerge constantly, requiring regular list updates. Some legitimate services or network configurations might use similar protocols or ports as proxies, leading to false positives. Proxy providers employ various techniques to mask their signatures, making detection more challenging. Some methods, like WebRTC or DNS leak detection, might raise privacy concerns if not implemented carefully. Also, port scanning is not straightforward, as an intermediate firewall can be easily configured to drop packets sent for port scanning. Similarly, the packets sent to identify the device's OS apart from the TCP connection can be quickly dropped using a properly configured firewall.

% Bot Detection
\subsubsection{\textbf{Bot Detection}} 
Bots are automated software programs that perform tasks online. While some bots are helpful, malicious bots can pose security threats. Bot detection identifies and differentiates between human users and automated bots \cite{bot, bot-method}.

\subsubsection*{Methodology} A few relevant methodologies in this regard are as follows.
        \begin{itemize}
            \item \textit{Behavioural Analysis:} Monitoring user activity patterns for signs of automation, such as inhumanly fast clicking or repetitive actions \cite{tax8, bot-accuracy, bot-method}.

            \item \textit{CAPTCHA Challenges:} Requiring users to solve challenges (like identifying images) that are difficult for bots but easy for humans \cite{tax8, bot-accuracy}.

            \item \textit{Device Fingerprinting:} Analysing device characteristics (like browser type, operating system) to identify patterns associated with known bots \cite{tax8}.

            \item \textit{IP Reputation Analysis:} Checking IP addresses against databases of known malicious bots \cite{tax8, bot-accuracy, bot-method}.
            
        \end{itemize}

\subsubsection*{Accuracy Concerns} Simple rule-based detection can be effective for identifying basic bots but less effective for advanced ones that adapt their behaviour. Analysing user behaviour patterns with machine learning can improve accuracy but may still have limitations \cite{bot, bot-accuracy}.

\textit{Limitations:} Sophisticated bots constantly adapt their behaviour to mimic human activity, making detection more challenging. Legitimate users with unusual browsing habits might be flagged as bots, leading to disruptions. Some techniques, like device fingerprinting, might raise privacy concerns if not implemented carefully. Advanced bots can utilise techniques like machine learning to solve CAPTCHAs, rendering them less effective \cite{bot}.

% Blocklist Check
\subsubsection{\textbf{Blocklist Check}} A blocklist check is a security technique used to identify and potentially block malicious or suspicious IP addresses, URLs, or email addresses. These blocklists are essentially large databases containing known threats compiled by security vendors, government agencies, or collaborative efforts \cite{blacklist-accuracy}.

\subsubsection*{Methodology} The IP address, URL, or email address in question is compared against entries in a blocklist database \cite{tax11, tax12, blacklist-method}.

\subsubsection*{Accuracy Concerns} Reputable, regularly updated blocklists offer higher accuracy but might still have occasional false positives. On the other hand, free or less-maintained blocklists have a higher chance of errors and outdated information, reducing their effectiveness \cite{blacklist-accuracy}.

\subsubsection*{Limitations} Blocklists may contain inaccurate or outdated information, leading to the blocking of legitimate traffic. New threats emerge constantly, and blocklists might not include them immediately, creating a window of vulnerability. Blocklists often focus on specific threat types, and a clean blocklist check does not guarantee complete security. Malicious actors can use techniques to avoid being blocklisted, like using constantly changing IP addresses. 

% Threat Detection
\subsubsection{\textbf{Threat Detection}} This method checks an IP address against databases of known malicious IPs associated with spam, malware, or other threats. This helps to identify potentially risky connections and implement mitigation strategies \cite{threat}.

\subsubsection*{Methodology} By checking databases which provide reputation details of IP addresses based on their behaviour and Internet history \cite{tax2, tax3}.

\subsubsection*{Accuracy Concerns} The accuracy of the reputation database itself is crucial. Regularly updated and well-maintained databases with reliable sources are more likely to provide accurate information. The frequency of database updates is essential. Outdated information can lead to ``false positives," where legitimate traffic gets blocked, or ``false negatives," where malicious activity goes undetected.
            
\subsubsection*{Limitations} Malicious actors can mask their actual IP address using techniques like spoofing. This can make it difficult for IP reputation analysis to identify the correct source of threats. The database's coverage of different geographical regions can impact accuracy. Some regions might have less comprehensive data available, leading to potential gaps in identifying threats originating from those areas. Organizations configuring IP reputation systems need to carefully set thresholds for blocking or flagging suspicious activity. Overly aggressive settings can lead to false positives, while overly lax settings might miss actual threats.

% Tools
\subsection{Tools} \label{tools}
% Port Scan
\subsubsection{\textbf{Port Scan}} Port scan is a technique used to identify active services (programs) running on a computer or network device \cite{port, port-method}.

\subsubsection*{Methodology} It works by sending specially crafted packets to different ports on the target device and analysing the response. Each port has a designated purpose (e.g., port 80 for web traffic, port 22 for Secure Shell access). \cite{port, port-method}

\subsubsection*{Accuracy Concerns} TCP SYN scan is a basic scan which checks for open ports by sending a connection request (SYN packet) and analysing the response. It is fast but might miss some services or be susceptible to evasion techniques like firewalls. Techniques like TCP scan, UDP scan, or stealth scans offer more information but might be slower or require more specialised tools. Also, they are not entirely immune to firewalls and similar protection methods. \cite{port}

\subsubsection*{Limitations} Port scanning only reveals open ports, not necessarily the specific vulnerabilities associated with those services. Malicious actors might use techniques to mask open ports or deflect scans, making them harder to detect. Scanning techniques may misinterpret responses or miss certain open ports depending on the configuration of the target device. Scanning large networks with many devices can be time-consuming and resource-intensive. \cite{port}

% \textit{Purposes:}
%         \begin{itemize}
%             \item  \textit{Identify vulnerabilities:} Open ports can reveal the types of services running on a device, which can then be checked against known vulnerabilities for potential security risks or to check whether the host is using some special kind of service like VPNs or proxies etc.

%             \item \textit{Inventory devices and services:} Scanning a network can help identify all connected devices and the services they offer, aiding in network management and security assessments.
    
%             \item \textit{Troubleshoot network issues:} Identifying open or closed ports can help diagnose connectivity problems or service disruptions.
            
%         \end{itemize}

% Whois
\subsubsection{\textbf{WHOIS}} The WHOIS (pronounced ``who is") service is a query and response protocol used to find information about a specific domain name registration \cite{whois, tax9, whois2}.

\subsubsection*{Methodology} When you enter a domain name in a WHOIS tool, it acts like a client, contacting the relevant server based on the domain extension (e.g., .com) and sending the name as a query. The server searches its database and sends back information like the registrant (depending on privacy settings), registration date, and nameservers. The tool then displays this information \cite{tax9}.
        
\subsubsection*{Accuracy Concerns} The information displayed might not always be wholly accurate or up-to-date, depending on the registrant's data entry \cite{whois2}.

\subsubsection*{Limitations} Many domain registrars offer privacy protection services that mask the registrant's contact information in WHOIS results. WHOIS does not reveal ownership details for some domain name extensions (like .gov or .edu). Also, the information might not always be accurate if not updated regularly \cite{tax9, whois2}.

% \textit{What it provides?:} \begin{itemize}
%             \item \textit{Registrant information:} Depending on privacy protection settings, WHOIS can reveal the name, organization, address, and contact details of the domain name registrant (the person or entity who registered the domain).

%             \item \textit{Domain registration details:} Information like the creation date, expiration date, and the domain registrar (the company that manages the domain name registration) can be obtained.

%             \item \textit{Nameserver information:} WHOIS can reveal the nameservers associated with the domain, which are the servers responsible for directing traffic to the website.
            
%         \end{itemize}

\begin{table*}[tb]
    \centering
    \begin{tabular}{|l|c|c|c|c|c|c|c|c|c|c|c|c|c|}
    \hline
        \textbf{Tool} & \textbf{\rotatebox{90}{Whois}} & \textbf{\rotatebox{90}{Geolocation}} & \textbf{\rotatebox{90}{Port Scan}} & \textbf{\rotatebox{90}{Liveness Test}} & \textbf{\rotatebox{90}{Proxy Detection}} & \textbf{\rotatebox{90}{Blocklist Check}} & \textbf{\rotatebox{90}{VPN Detection}} & \textbf{\rotatebox{90}{Tor Detection}} & \textbf{\rotatebox{90}{Bot Detection}} & \textbf{\rotatebox{90}{Threat Detection}} & \textbf{\rotatebox{90}{IPv6?}} & 
        \textbf{\rotatebox{90}{Web App}} & 
        \textbf{\rotatebox{90}{API}} \\
        \hline
        \url{https://www.ipqualityscore.com} & & F & & & F & F & F & F & F & F & TRUE & F & F\\
        \hline
        \url{https://ipinfo.io} & F & F & F & F	& F	& & F & F	& & F & FALSE & F & F \\
        \hline
        \url{https://www.neutrinoapi.com} & & F & & & F & F & F & F & F & F & TRUE & F & F\\
        \hline
        \url{https://ipapi.com} & & F & F & F & F & & & F & F & & TRUE & F & F \\ \hline
        \url{https://www.maxmind.com/en/geoip-databases} &  & P & & & P & & & & & & TRUE & P & P \\ 
        \hline
        \hline
        \url{https://ipanalyzer.in} (Our tool \cite{IPANALYZER}) & F & F & F & F & F & F & F & F & F & F & TRUE & F & F \\
        \hline
        %\vdots & \vdots & \vdots & \vdots & \vdots & 
        %\vdots & \vdots & \vdots & \vdots & \vdots & 
        %\vdots & \vdots & \vdots & \vdots \\ \hline 
    \end{tabular}
    \caption{The table shows 5 out of 105 surveyed tools. F means free feature, P means paid feature whereas blank means feature is missing. For a complete survey, please check \cite{survey}.}
    \label{tab:survey}
\end{table*}

\begin{figure*}[bt]
	\centering
        \includegraphics[width=\textwidth]{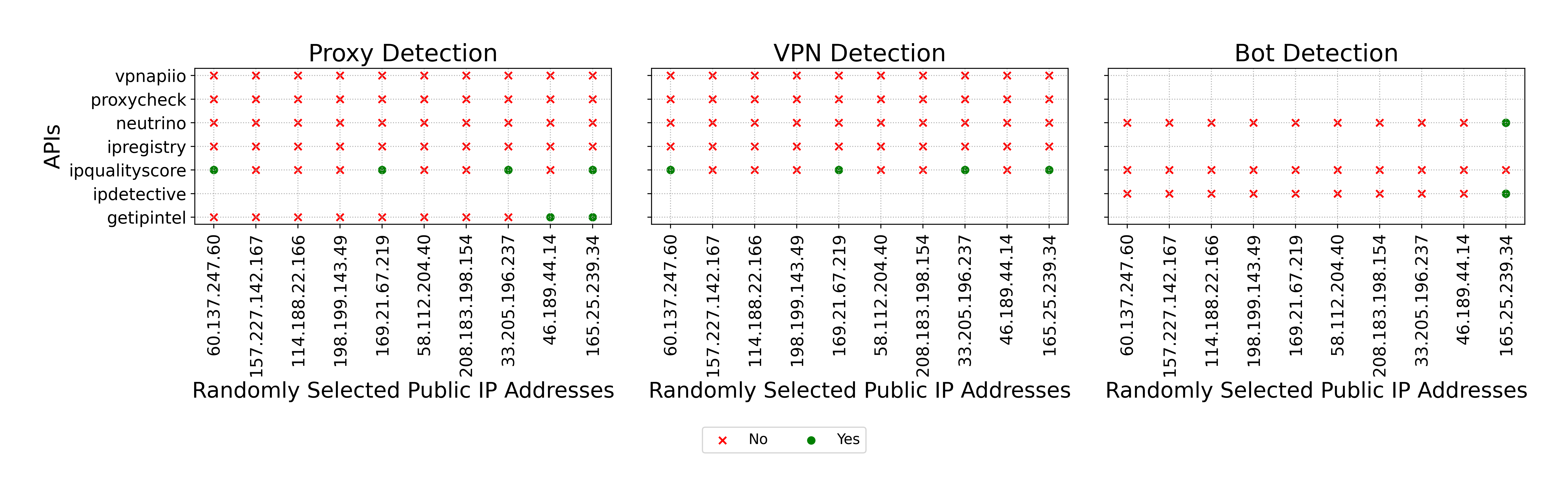}
        
    \caption{Output of API Calls from different web tools of 10 random public IP addresses for Proxy, VPN, and Bot Detection.}
    \label{results}
\end{figure*}

% Liveness Check
\subsubsection{\textbf{Liveness Check (Ping Test)}} The ping service, also referred as ``pinging," acts like a digital echolocation tool for networks. Ping measures the round-trip time (RTT) – the total time it takes for the packet to travel from your device, reach the destination, and return with a response \cite{ping}.

\subsubsection*{Methodology} It operates by sending a small data packet, like a digital ping, from your device to a specific destination (identified by an IP address) on the Internet. This destination could be a website, another computer, or any device connected to the Internet. Once the packet reaches the target, it sends a response back to your device \cite{tax13, ping}.

\subsubsection*{Accuracy Concerns} High network traffic can affect the RTT value, potentially making it appear higher than the actual responsiveness. This can lead to misinterpretations of network performance. Firewalls or other security policies might filter out ping requests, resulting in inaccurate results or no responses. The absence of a response from the host can sometimes be misinterpreted as the host not being live, even though that might not be true \cite{ping-accuracy}.

\subsubsection*{Limitations} Ping measures the round-trip time and does not provide detailed information about the network path, potential bottlenecks, or the cause of connectivity issues. Ping only tests the reachability of the target device or service based on the provided IP address. It does not delve deeper into the internal health or functionality of the target system or application \cite{ping-accuracy}.

% Related Work
\section{Related Work}
In this section, we present a comprehensive survey of various tools related to IP address analysis. \cite{survey} presents the complete survey, while Table~\ref{tab:survey} provides its snippet. Each row represents a specific tool, while columns represent distinct features. Within this framework, the designation ``F" signifies features included at no additional cost, ``P" denotes features requiring a premium subscription, and blank cells indicate a feature's absence.

Several key observations can be drawn from this systematic evaluation. Geolocation and WHOIS lookup functionalities are identified as the most common, with little to no deviation in results observed across the analysed tools. Conversely, features like Bot detection, Tor detection, and Threat Detection are encountered less frequently. Interestingly, no single tool provides a complete suite of all evaluated features.

Furthermore, the analysis reveals that a significant majority (96.36\%) of tools function solely through web interfaces, while a minority (21.82\%) offer access through Application Programming Interfaces (APIs). Notably, a substantial proportion (96.07\%) of features are offered free of charge, albeit with potential limitations on usage. It is essential to acknowledge that much deviation in results is observed across multiple tools analysed for features like Threat Detection, VPN detection, and IP Geolocation. These variations likely stem from discrepancies in the underlying data sources employed.

Figure \ref{results} showcases a few test results. For ten randomly selected public IP addresses across seven IP address analysis web tools, we check the existence of three features in them -- namely, whether each of them is behind a proxy, VPN or is a bot. A \mydot[green] represents \textit{presence}, whereas a \mycross[red] represents \textit{absence} of the feature tested by the respective web tool for the respective IP address.

These observations highlight the complex nature of the tool landscape and shed light on potential factors influencing discrepancies in analysis outcomes, particularly for features like IP reputation and geolocation.

\section{Proposed Solution}\label{sec:proposed}

In this section, we present a comprehensive tool which has been developed using Python~\cite{python}, Angular framework~\cite{angular} for frontend and Django web framework for the backend~\cite{django}. Our tool provides robust user management with a suite of functionalities crucial to IP address analysis. On top of that, our tool implements various measures to counteract various threats, including but not limited to Cross-Site Request Forgery (CSRF) \cite{csrf} attacks, Cross-Site Scripting (XSS) \cite{xss} and JavaScript Injection Attacks \cite{javascript}. In addition, the two-factor authentication feature is included in the solution to make the application more secure. Moreover, our tool is designed to provide an intuitive user experience, even for users with limited technical expertise.

Our tool \cite{IPANALYZER} focuses on delivering functionalities for the comprehensive analysis of IP addresses. %Among essential features is IP geolocation, which allows users to find out what geolocation is linked to a specific IP address. 
Our application queries several databases through APIs to detect anonymisation services like proxies, VPNs and the Tor network, which people use to disguise actual IP addresses. The tool can also identify whether IP is linked to a human or a bot. Moreover, our tool can detect whether that bot was involved in malicious activities by utilising the history of that IP address in databases.
Furthermore, our tool has an option for port scans. While scanning an IP address, by analysing particular ports, it finds out the open services. Integration with WHOIS databases provides easy access to the publicly available registration information associated with the domain name of interest. Additionally, our tool can verify whether a specific IP address is listed in any blocklist known to maintain a list of malicious IP addresses so that the user may evaluate the dangers of encountering the IP address.

In addition to this, our tool computes a confidence score for each of the outputs based on the API responses received. Our tool uses several APIs that act as gateways to various global databases. This makes it possible to extract key data points. After receiving the responses for each functionality requested, our tool computes the weighted average of the responses based on predefined weights to give a confidence score to each of the predicted outputs.

Operating as a web application, our tool can be deployed on a single server to be accessible over the network from any point. It can be accessed globally if our tool is connected to the internet. In a private network, a predefined set of users can access the tool using the user management system it incorporates, which is usually beneficial for corporate and academic institutions. As we have used the Angular framework for our front end, we are able to achieve a high level of responsiveness, while on the server side, we use Django, ensuring scalability. 

In addition, to provide a second opinion for the purpose of verifying the results obtained, there is also an option to compute the abuse score of a particular IP address. For this purpose, we use the AbuseIPDB API~\cite{abuseipdb} to fetch user reports and abuse scores for an IP address or domain name, which are elegantly incorporated into the web interface of our tool. One of the appealing features of AbuseIPDB is the vast community of people who provide reports for IP addresses. Our tool integrates user reports from the past 90 days for an IP address through AbduseIPDB, including the report time, report place, report categories~\cite{report-cat}, and comments. Only registered users can add reports to an IP address in the AbuseIPDB. Assigned abuse score to an IP address is defined based on user reports, with each report being weighted according to the respective user's weight. Although the formula for weighted sum and abuse score calculation is not clearly mentioned in the AbuseIPDB documentation, the weighted user scoring system ensures that no single user can highly influence the abuse score of an IP. The abuse score is provided on a scale of 0 to 100, with higher scores meaning the IP address is more likely to be associated with malicious activity. The API also indicates whether the given IP address belongs to a Tor network or an Internet service provider.

Finally, our tool includes a local database that improves its performance and allows offline usage. For this, we have chosen MongoDB~\cite{mongodb}, ascribing to its scalability and resilience. It saves the users' records along with the logs of searched IP addresses and domain names with their information. In case of poor API communication or to save API calls, a user can choose to get the latest information saved in the logs, as it is less likely that the information would have changed in that short period of time. This ensures the smooth functioning of the system and facilitates informed decision-making, even with poor API connectivity. %For reference, please see Figure \ref{tool} for some images of our tool.

\section{Conclusion and Future Scope}
%The objective of the present work is to conduct a review of the available options that can be used to study IP addresses and provide a feature-based comparison and statistical evaluation of the different tools. 
In this work, we have conducted a thorough review of the public and proprietary tools available to analyse and characterize the IP addresses and presented  feature-based comparison and statistical evaluation of different tools.
Our work tries to explore the rarity of a specific set of features and how challenging they are to implement. 
In addition, we have developed and proposed the IP analysis tool, which has a well-balanced range of functionalities and is backed by a modern architecture and a superior user management system. However, there is ample scope to enhance the features and transparency of the tool. %can be improved further.

Although the current instrument we employ is useful, there are many other features that could enhance its range and depth. Employing a historical database of registered domain names would, for instance, allow the users to see the changes in ownership and registration over a given period for a particular IP address. In addition, the growing advancement of the network makes it imperative to include support for IPv6 – a protocol with additional networking that enables the assessment of more internet devices and networks. The introduction of visualisation tools in the report generation process will significantly improve the user experience because they will be able to see trends and patterns more quickly. In order to enhance security management practices, an automated notification system that measures certain user-specified conditions and informs the user whenever a certain IP or domain becomes available should be put in place.

Overall, these changes would enhance the tool’s flexibility and would further strengthen its reputation as an advanced and reliable IP analysis tool.

\bibliographystyle{IEEEtran}
\balance
\bibliography{references}
% \printbibliography

%Link to the survey table: \url{https://docs.google.com/spreadsheets/d/1tQcEFkUygEJbKDMkUVCawtYDMy9Ydo-X/edit?usp=sharing&ouid=100824980323805030181&rtpof=true&sd=true}.

\end{document}